\documentclass[runningheads,a4paper]{llncs}

\usepackage{amssymb}
\usepackage{graphicx}

\usepackage{url}

\usepackage{amsmath}
\usepackage{tikz}
\usetikzlibrary{calc}

\usepackage{caption,subcaption}
\usepackage{multirow}

\usepackage{booktabs}

\usepackage{epstopdf}

\usepackage{listings}

\author{Marek Kokot \and Sebastian Deorowicz \and Maciej D{\l}ugosz}
\authorrunning{Marek Kokot \and Sebastian Deorowicz \and Maciej D{\l}ugosz}
\institute{Institute of Informatics, Silesian University of Technology,\\
Akademicka 16, 44-100 Gliwice, Poland\\
\path|{marek.kokot,sebastian.deorowicz,maciej.dlugosz}@polsl.pl|}

\title{Even faster sorting of (not only) integers}

\begin{document}
\maketitle

\begin{abstract}
In this paper we introduce RADULS2, the fastest parallel sorter based on radix algorithm.
It is optimized to process huge amounts of data making use of modern multicore CPUs.
The main novelties include: extremely optimized algorithm for handling tiny arrays (up to about a hundred of records) that could appear even billions times as subproblems to handle and improved processing of larger subarrays with better use of non-temporal memory stores.
\end{abstract}

\section{Introduction}

Sorting is one of the fundamental tasks in computer science. Despite the fact that the problem was discussed many times from both theoretical and practical perspective, it remains open and improved solutions are still proposed.
The main reason of its popularity is a wide range of applications.
Frequently sorting is an important component of other algorithms.
Ordered sequences may often be faster processed or stored in compacted form. 
Sorting algorithms may be divided into two groups: comparison based and non-comparison based. 
In the first group, the elements of input sequence are compared in pairs, which implies the lower bound of the number of comparisons $\Omega(n\log{}n)$~\cite{ref:Knu1968}. 
In the second group, no comparisons are made, thus the lower bound may be relaxed to $\Omega(n)$.
Unfortunately, non-comparison attempt requires additional assumptions about the data, which makes this group of algorithms less general. 
Despite the algorithm category, the technical details of implementation have crucial meaning in practice.\looseness=-1

One of such aspects is memory access pattern. 
In modern computer architectures, processors are equipped with small amount of fast cache memory, which minimizes the negative impact of high latency of the main memory access.
The cache is usually divided into a couple of levels: L1, L2, L3. 
When a processor core needs to load a particular memory location, it is first checked if the required address is already in the cache (checking starts from the lowest level and continues to the highest one). 
If the required address is absent in the cache it must be loaded from the main memory. Such a situation is called a {\it cache miss} \cite{ref:dre2007}. Algorithms should be constructed to avoid cache misses whenever possible.

The second important factor on the way to fast algorithms is parallelization. Nowadays even desktop computers and laptops are equipped with multicore processors, which may be used to increase performance. In practice only for simple algorithms optimal speedup is easily achievable. 
For the complex ones it is often hard to keep the desired speedup when the number of involved cores increases. 
Combining both cache-friendly memory access and parallelization introduces new issues that need to be addressed. In a multicore system each core has its own L1 cache, while L2 may be assigned to a single core or may be shared among a group of cores. L3 is typically common to all cores in a chip. The data in the cache is organized in lines (typically of 64 bytes). When two cores operate on a separate part of the same cache line and at least one of them performs writing, the slow synchronization occurs. Such a situation is called {\it false sharing}~\cite{ref:Int2011} and similarly as cache misses should be avoided.
 
The other important factor is understanding the cost of {\it branch mispredictions}~\cite{ref:fog2011}. Modern processors are equipped with long pipelines \cite{ref:she2013}. For each conditional instruction processor predicts how it will be evaluated and chooses the next instructions to load into pipeline. If the actual result of comparison is different than predicted then branch misprediction occurs and the whole pipeline must be cleaned, which introduces long delay. In some cases it is possible to avoid branch mispredictions using conditional assembly instructions, e.g., {\it CMOVcc}~\cite{ref:Ede2016}.

Radix sort \cite{ref:cor2009} is a non-comparison technique and has the time complexity  $\mathcal{O}(kn)$, where $n$ is the number of elements to be sorted and~$k$ is the length of keys in digits. The digit is a fixed-size part of a key, e.g., in case of sorting strings a single character. In general radix sort works in $k$ phases, each phase is related to sorting keys according to the current digit. Digits are usually from a small alphabet and counting sort may be applied per single phase. There are two categories of radix sort depending on the order in which digits are processed. The first variant, LSD (least significant digit), starts sorting whole data from the least significant digit and continues to the most significant one. It is required to use a stable sort per each phase to ensure that the final result is valid. The second variant, MSD (most significant digit), works in the opposite direction and starts sorting the data according to the most significant digit. This results in partitioning data into bins that may be further processed independently in the recursive manner.\looseness=-1

The MSD radix has an important advantage over the LSD variant. 
The bins created after each phase tend to shrink (except bins with all keys equal).
After a few phases the bins become tiny and switching to other algorithm improves the overall performance. 
A common approach here is to use some comparison-based sorting algorithm, e.g., insertion sort~\cite{ref:Mcl1993}.
Typically, the time of handling such tiny bins could be from 10 to 50 percent of total time of radix sort.

Sorting tiny arrays is a special case, which should be considered separately. Often algorithms with poor time complexity perform well for tiny arrays.
Therefore many complex sorting algorithms, like introspective sort \cite{ref:Mus1997}, uses them as subroutines. Yet tiny array sorting seems to be not fully investigated. 


Recently we proposed RADULS \cite{ref:Kokot2016}, the fastest existing radix sort.
In this paper we present its significantly improved successor.
The main novelties in RADULS2 are motivated by the following investigations. 
Firstly, we considered the special case of tiny array sorting.
We proposed a new hybrid algorithm, which is significantly faster than the hybrid of insertion sort, Shell sort, and introspective sort used in RADULS. It also outperforms insertion sort used in common std::sort function implementations in the C++ standard libraries.
Secondly, we optimized the last phase of radix sort, in which the subarrays are small, but not so small to satisfy the tiny bin criterion.
Thirdly, we improved the processing of small bins to take the most of non-temporal memory stores present at modern CPUs.\looseness=-1

The paper is organized as follows. In Section~\ref{sec:related} we give a short description of existing radix sorters and algorithms that are capable to sort tiny arrays efficiently. Our algorithm is described in Section~\ref{sec:algorithm}. The experimental results are presented in Section~\ref{sec:results}. 
In the last section we conclude the paper.

\section{Related works}
\label{sec:related}
\subsection{Radix sorters}
Probably the fastest version of LSD radix sort was proposed by Satish \emph{et al.}~\cite{ref:Sat2010}.
The main idea is to divide the scatter phase into two steps to make it cache-friendly. 
Before saving to the main memory, each record is stored into a cache buffer. 
When the buffer for some digit value becomes full, it is transfered using a non-temporal memory store to the main memory. 
The size of buffer per single digit value is a multiple of 64\,B (size of a~single memory transfer). This approach increases bandwidth utilization and reduces number of cache misses.

In \cite{ref:Cho2015} an efficient MSD radix sort was proposed. Its in-place nature results in difficulty of parallelization. 
The algorithm performs series of element swaps to sort them according to the current digit.
Unfortunately, there are high dependencies between the operations, which make them hard to perform in parallel. 
The authors introduced {\it speculative permutation} and {\it repair} phases to address this problem. Another issue occurring is significant imbalance of bins sizes.
The authors handles this by a {\it distribution-adaptive load balancing}.\looseness=-1

In \cite{ref:Kokot2016} we proposed variant of MSD radix sort. The main idea is to perform appropriate optimization depending on the current phase and size of currently processed bin. For instance, if the current bin is big enough, a buffered version of scattering inspired by Satish \emph{et al.} is used. 
When the current bin is tiny (say contains tens of elements), some comparison-based sorting algorithm is chosen. 


\subsection{Tiny array sorters}
Sorting tiny arrays is not as widely discussed in the literature as the case of big ones. For most applications it is irrelevant which sorting algorithm is chosen if the number of records is low. Still many sorters stops their default procedure and switch to a simpler algorithm when the number of remaining records is below given threshold. 
Such a situation may occur billions of times for large inputs. 
Thus, improving tiny sorting procedure will also improve overall running time of given sorting algorithm. 
For example introspective sort \cite{ref:Mus1997}, which is usually a part of standard C++ library is a hybrid of quicksort \cite{ref:Hoa1962} and heapsort \cite{ref:Wil1964}, yet for tiny ranges it switches to insertion sort.\looseness=-1

In the case of sufficiently tiny arrays all the data are usually in L1 cache, so there are no cache misses, and the cost of memory accesses is low.
Thus, the dominant factors that have impact on the running time are:
(\emph{i}) simplicity of the algorithm, that usually is reflected in the small number of jump operations,
(\emph{ii}) number of comparisons, and the cost of single comparison,
(\emph{iii}) number of assignments.
As the assignment is usually a simple (branchless) operation  it is faster than comparison, which, especially for longer than 8\,B keys, are hard to be made without jumps.

As said above, insertion sort is usually taken as a ``tiny sorter''.
Shell sort~\cite{ref:She1959} being its generalization is also an obvious candidate.
%
Sorting networks \cite{ref:Bat1968} are constructed with series of so-called {\it comparison elements} or {\it comparators}. 
Each such element performs a conditional swap of its arguments. 
The number of such operations, as well as the indices of the compared records, are fixed for a given input size, i.e., they are independent on the ordering of input elements. 
This allows the compiler to hard-code the indices in the code, so the addressing is faster than when the index must be loaded from a variable.
There are two criteria of sorting-network optimality. The first one is related to its {\it depth} defined as the number of layers of comparators (a layer is a set of comparators which may perform operations independently). The second criteria is sorting network {\it size} defined as the total numer of comparators. Optimal-depth sorting network are known only up to 17 records \cite{ref:Ehl2015}, optimal-size networks are known only up to 10~records \cite{ref:Cod2014}.\looseness=-1

In the {enumeration sorting}~\cite{ref:Knu1968} the final position of each element is determined by comparing it against each other. 
The number of comparisons is $\mathcal{O}(n^2)$, but the number of element moves is only $\mathcal{O}(n)$. 
Even if this algorithm requires high number of comparisons, they may be implemented branchless with {\it SETcc} assembly instruction. It allows casting boolean comparison result to integer value, which may be further used to update final position of the element.





\section{Our algorithm}
\label{sec:algorithm}
\subsection{General idea}
Our algorithm is an MSD radix sort with 1\,B digits. 
It follows the same general scheme as its predecessor~\cite{ref:Kokot2016}:
\begin{enumerate}
\item Parallel processing of the whole array according to the first digit in a cache-efficient way.
\item Serial processing of small bins obtained in the previous stages, in a cache-efficient way. Several bins are processed by separate threads in the same time.\looseness=-1
\item Serial processing of tiny bins (containing up to tens or hundreds of records) using a hybrid of insertion sort, Shell sort, introspective sort. Several threads process different bins in parallel.
\end{enumerate}

The main improvements in this paper are in the small and tiny bin processing subroutines.
These phases constitute the majority of RADULS processing time, so the impact of the proposed novelties on the total sorting time is remarkable.

%


\subsection{Small bin processing}
Small bins are processed in a decreasing order of their sizes. 
Each thread handles a single bin at a time. 
There are two cases here. 
In the first one, the bin is small enough to fit in a half of L2 cache, so cache misses are unlikely.
In such a situation simple counting and scatter approach is chosen.

Larger bins will likely do not fit into cache, so many cache misses can be expected.
Thus, a buffered version similar to the first phase (motivated by Satish~\emph{et al.}~\cite{ref:Sat2010})  is chosen.
Nevertheless, there are some traps here.
In the first pass it can be assured that the input array is aligned to 64\,B which allows to use efficient non-temporal memory storage operations. 
In the following stages, the beginning address of subarray representing bin is likely unaligned.
Thus, non-temporal memory storage operations used in RADULS are inefficient here.
Depending on the CPU architecture they are a few times slower than in the case of aligned address.
(Usually, the difference is larger for AMD Opteron than Intel Xeon.)\looseness=-1

To solve this problem in RADULS2 we implemented special alignment of subarrays.
The beginning of the aligned bin address can be localed in other bin.
Nevertheless, the transfers from buffers (located in cache) to the main memory are made in such a way that other bins are not overwritten.
Such solution resulted in about two times faster scatter step than in RADULS.

\subsection{Tiny bin processing}
Taking into account the results of our experiments on ``tiny sorters'' (discussed in Section~\ref{sec:results}) we constructed a hybrid algorithm that is able to handle tiny arrays efficiently. 
The algorithm exploits advantages of sorting networks (SN)~\cite{ref:Bat1968} and enumeration sort (ES) \cite{ref:Knu1968} for the tiniest arrays, while for larger ones it uses our modification of BlockQuicksort (BQS)~\cite{ref:Ede2016} in which we replaced insertion sort for the tiniest arrays by sorting network or enumeration sort. 

For the smallest records (i.e., containing just 8\,B key) the hybrid switches at $n=64$ elements from  SN to BQS with SN for the tiniest arrays.
The sorting networks for $n>64$ seem to be of almost the same speed as BQS, but the source code for SN becomes long (and also the code must be prepared for each array size), which leads to long compilation and large executable.

If the data field exists the situation becomes more complex as conditional swaps of records in the array takes longer time, and often cannot be made without jumps (or at least compilers tend to prepare the code with branches).
Therefore, ES becomes more attractive, as it makes just $n$ (branchless) assignments.
Even the fact that ES makes $\mathcal{O}(n^2)$ comparisons seems to be acceptable (at least for tiniest arrays) if we take into account that the comparisons are branchless.
Therefore, for 16\,B or 24\,B data fields the hybrid employs ES when the number of elements is not larger than 24 and BQS with ES for the tiniest subarrays otherwise.\looseness=-1

The most complex situation is the case of 8\,B data field.
For $n\le 16$ the hybrid employs the ES, but then, up to 32 records, the SN is used.
It seems that the quadratic number of comparisons of ES becomes the problem for larger arrays.
For $n>32$ BQS with SN is used.

The situation of 16\,B keys and no additional data is different as it is hard to implement the comparison of elements in a branchless way.
Therefore, the proposed hybrid uses ES according to the most significant 8\,B word (to make use of branchless comparisons) and then fix the ordering using IS.
For $n>24$ we, however, employ BQS with the above ES+IS as a dispatch for tiniest subarrays.

%

In our general radix procedure we analyze each newly created bin to make a decision about its further processing. There are bins that do not meet the requirements of being tiny, but our general radix procedure introduce observable overhead due to unnecessary analysis of empty subbins. To address this problem we treat bins of size at most 256 in a special manner. 
During the histogram construction we store the identifiers of subbins containing at least 2 elements that need further processing.
Thus, the other subbins can be easily omitted.

\section{Experimental results}
\label{sec:results}
\subsection{Tiny array sorting algorithms}

All tiny sorters were implemented in the C++14 programing language. A couple of SSE instructions were used to improve the performance of sorting network in case of 16\,B records. For compilation we used GCC 6.3.0 and Visual C++ 2015. The experiments were run on four platforms: 
\begin{itemize}
\item Laptop with Intel Core i7-4700MQ CPU clocked at 2.4\,GHz (Windows),
\item Workstation with Intel Core i7-5820K CPU clocked at 3.30\,GHz (Linux),
\item Workstation with two Intel Xeon E5-2670v3 CPUs clocked at 2.3\,GHz (Linux),
\item Server with four AMD Opteron 6320 CPUs clocked at 2.8\,GHz (Linux).
\end{itemize}

\begin{figure}[p]
	\begin{center}
		\begin{tabular}{ccc}
			\includegraphics[width=0.48\textwidth]{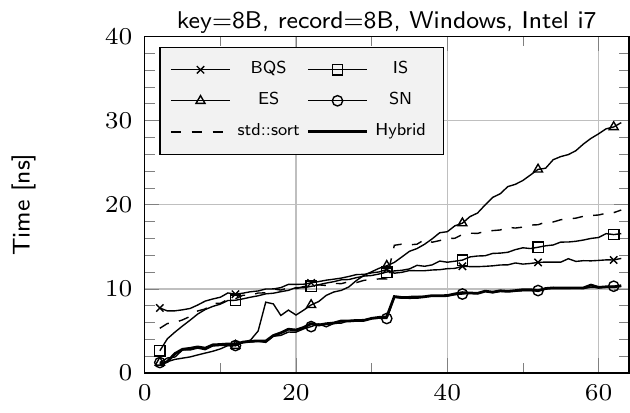} && 
			\includegraphics[width=0.41\textwidth]{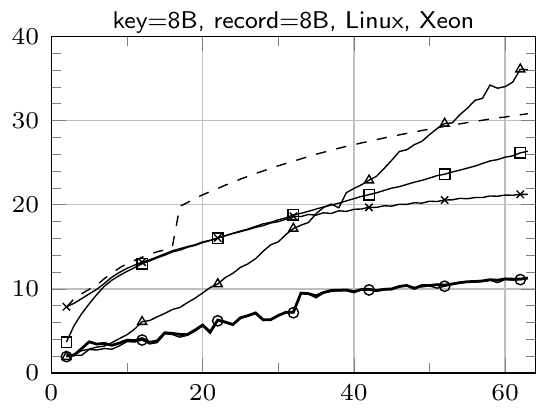} \\
			\includegraphics[width=0.48\textwidth]{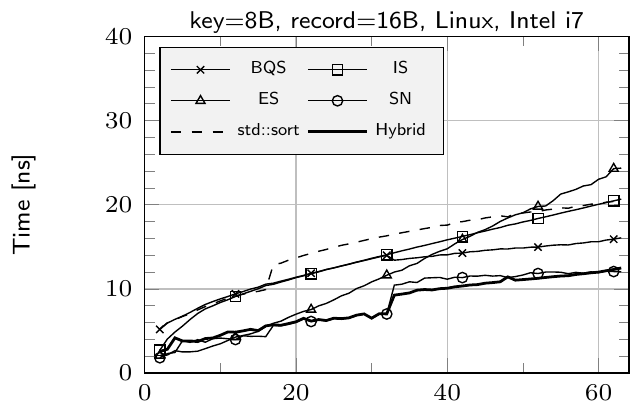} && 
			\includegraphics[width=0.41\textwidth]{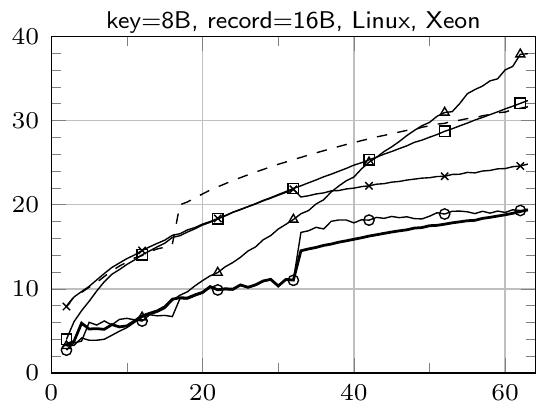} \\
			\includegraphics[width=0.48\textwidth]{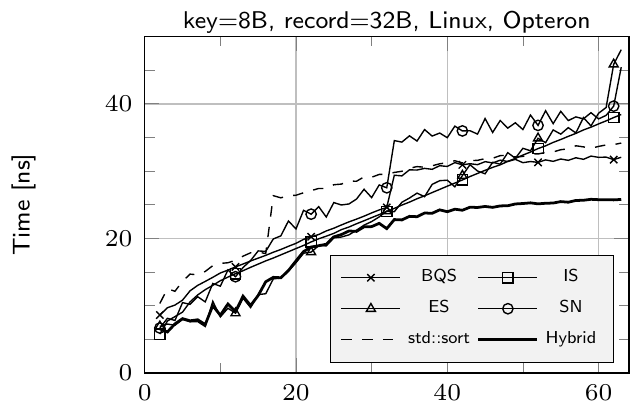} && 
			\includegraphics[width=0.41\textwidth]{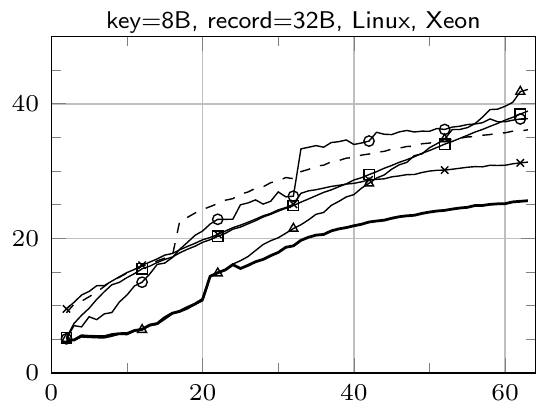} \\
			\includegraphics[width=0.48\textwidth]{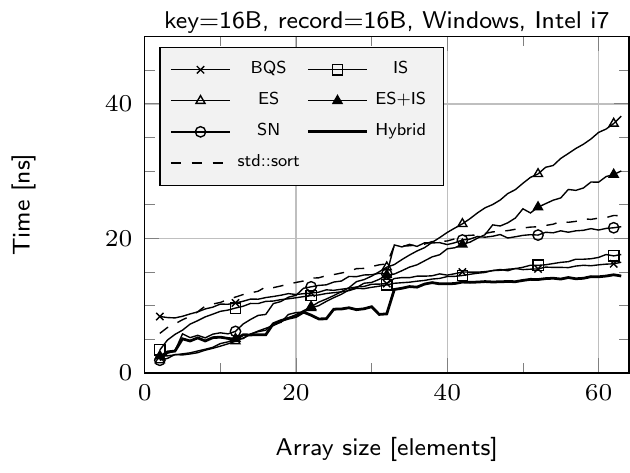} && 
			\includegraphics[width=0.41\textwidth]{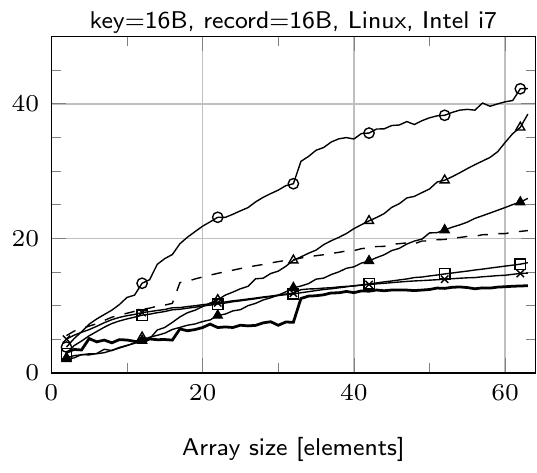} \\
		\end{tabular}
	\end{center}
	\caption{Running times (per element) for tiny sorters. 
The examined algorithms: BlockQuicksort (BQS), insertion sort (IS), enumeration sort (ES), sorting networks (SN), introspective sort from C++ (std::sort), enumeration sort according to most significant 8\,B word followed by insertion sort (ES+IS), our hybrid algorithm (Hybrid).
The legends are the same for both charts of each row.}
	\label{fig:sizes}
\end{figure}

The results of sorting tiny arrays of size from 2 to 64 records with the following algorithms:
insertion sort (IS), BlockQuicksort (BQS), enumeration sort (ES), introspective sort implemented as a part of standard C++ library (std::sort), sorting network (SN), our hybrid algorithm (Hybrid), enumeration sort followed by insertion sort (ES+IS) are shown in Fig.~\ref{fig:sizes}.
The running times are divided by the size of the sorted array.

The fastest method for records with 8\,B key and no data field or 8\,B data field is SN (except for the smallest number of records---up to 10). Yet it is worth mentioning, that our implementation of sorting network for 16\,B records uses SSE to allow elements swapping by branchless code. Without such optimization SN performed much worse as compilers tend to produce the code with branching instructions.
The cost of swapping the elements is easy observable in case of 32\,B records---in such a situation ES clearly outperforms SN.

The case of 16\,B keys and no additional data is quite interesting. The comparison of elements cannot be performed in a single step, thereby maintaining branchless code is harder. Our implementation of SN suffers in such situation, yet it is observable that Visual C++ is able to perform better optimizations, which leads to much lower runtime than in case of GCC. Our two-stage technique helps to remove branch mispredictions overhead, in fact our hybrid algorithm in such situation is SN according to 8\,B key and 8\,B data field followed by IS.\looseness=-1

The sudden rise of running time of SN above 32 records is an effect of much higher number of comparison elements. For such values we used Batcher's Odd-Even algorithm~\cite{ref:Bat1968} to generate networks. For smaller values we was able to use more efficient networks. In case of std::sort similar behavior may be explained by a threshold of switching to insertion sort.

\subsection{RADULS 2}
\begin{figure}[p]
\begin{center}
\begin{tabular}{lcr}
\includegraphics[height=1.7in]{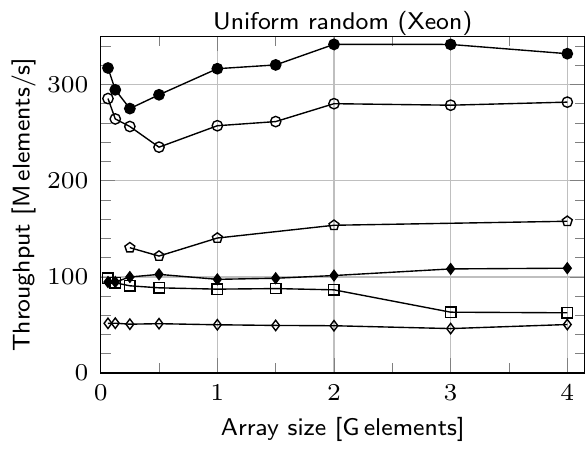} &&
\includegraphics[height=1.7in]{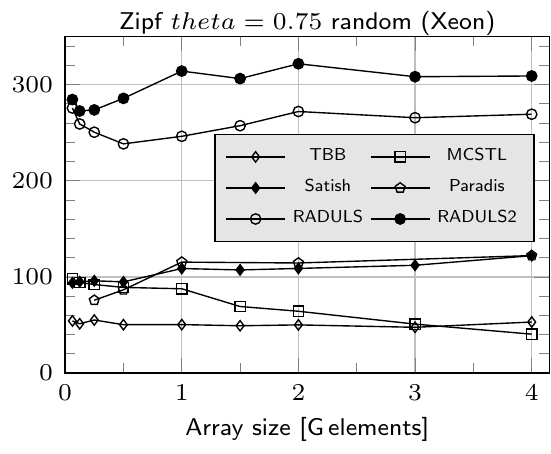}\\
\includegraphics[height=1.7in]{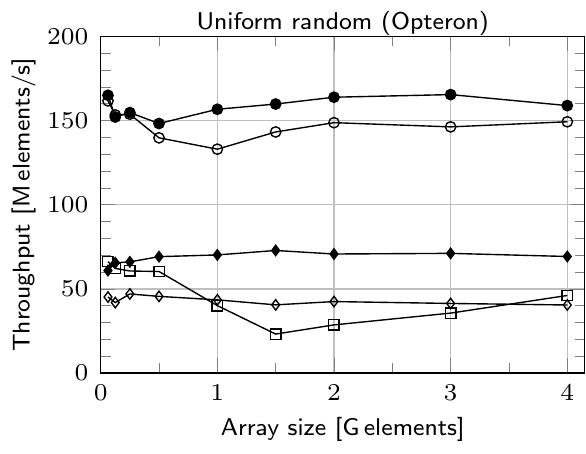} &&
\includegraphics[height=1.7in]{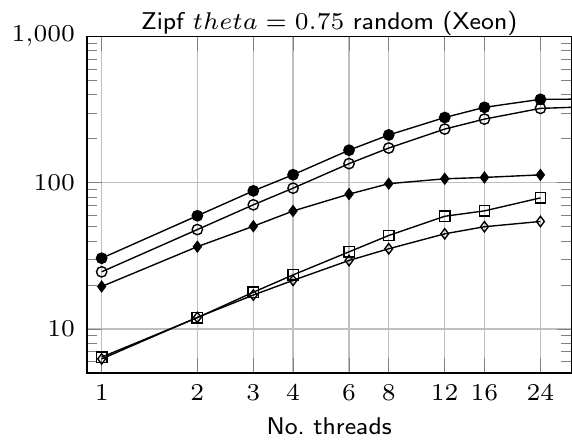}\\
\end{tabular}
\end{center}
\caption{Experimental comparison of best sorting algorithms.
The right bottom chart shows how the algorithm speed ups for growing number of threads.
The remaining charts present throughput for 16 threads when sorting 2\,G records of size 16\,B (8\,B key and 8\,B data).}
\label{fig:raduls}
\end{figure}

The results of experiments for RADULS 2 and the fastest competitors are given in Fig.~\ref{fig:raduls}.
Similarly as in~\cite{ref:Kokot2016} we picked for comparison the following algorithms:
\begin{itemize}
\item TBB---the parallel comparison sort implemented in the Intel Threading Building Blocks~\cite{ref:TBB} (2017 Update 3 release),
\item MCSTL---the parallel hybrid sort~\cite{ref:Sin2007}, included in GNU's libstdc++ library,
\item Satish---the buffered LSD radix sort introduced by Satish \emph{et al.}~\cite{ref:Sat2010} with the buffer size for a specific digit equal to 4 times the cache line size,
\item PARADIS---the state-of-the-art in-place radix sort algorithm by Cho \emph{et al.}~\cite{ref:Cho2015},
\item RADULS---the state-of-the-art radix sort algorithm~\cite{ref:Kokot2016}. 
\end{itemize}

The main experiments are for 16 threads for 16\,B records (8\,B key and 8\,B data).
As we can see RADULS2 is the fastest sorting algorithm in all the experiments. 
For 1\,G elements it is 23\% (uniform distribution) and 28\% (Zipf distribution) faster than RADULS when run at Xeon-based workstation.
At Opteron-based server the advantage is 18\%.
The scalability experiments show that RADULS2 scales almost in the same way as RADULS.

\begin{figure}[p]
\begin{center}
\includegraphics[width=0.98\textwidth]{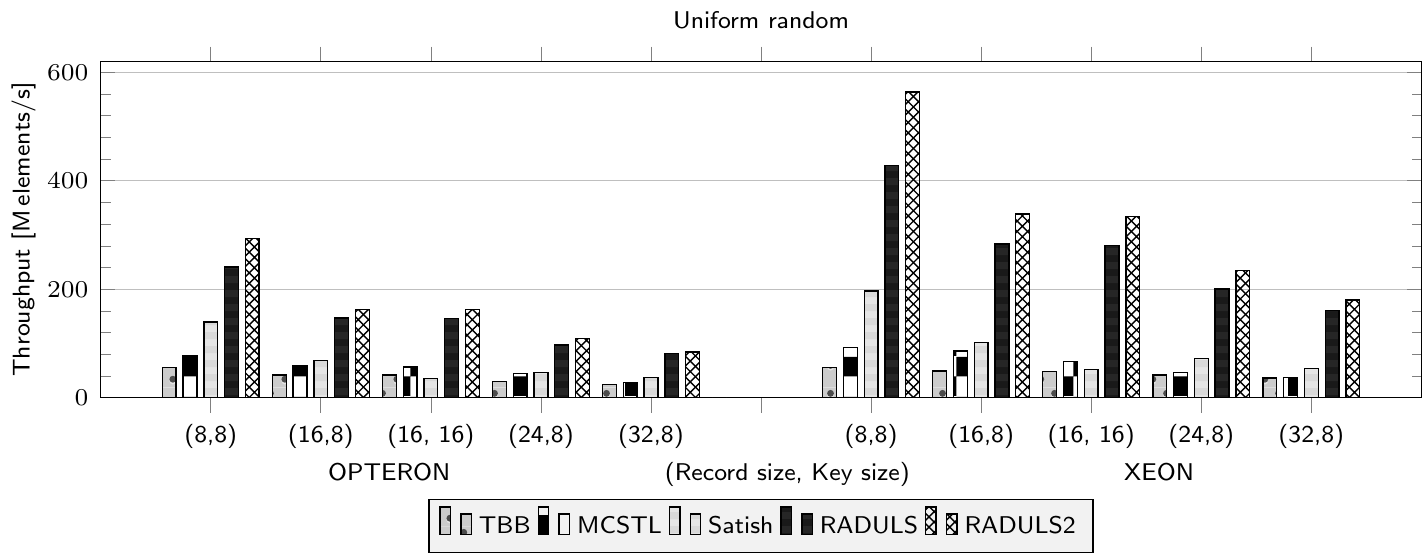}
\end{center}
\caption{Experimental comparison of algorithms when sorting 2\,G records of various sizes using 16 threads.}
\label{fig:records}
\end{figure}

We also evaluated the behavior of RADULS2 for records of different sizes (Fig.~\ref{fig:records}).
RADULS2 was noticeably faster than its predecessor, however, the difference becomes smaller for longer records.
It is also interesting to compare the relative throughput for different platforms.
Opterons were clocked higher (2.8\,GHz) than Xeons (2.3\,GHz) but sorted the data about two times slower.\looseness=-1

RADULS is used in KMC3~\cite{ref:Kokot2017}, our $k$-mer counter for genomic data.
The excellent performance of KMC3 is in significant part due to fast sorter.
When sorting KMC set from~\cite{ref:Kokot2016} the gain in throughput was 17\% (413\,M 16\,B-key records) and 10\% (1887\,M 16\,B-key records).

\section{Conclusions}
In this paper we examined sorting algorithms for tiny arrays, as well as we proved the relevance of this topic by applying proposed hybrid solution to RADULS---our radix sort algorithm. We also proposed other improvements, which---together with ``tiny sorters''---constitute new version, i.e., RADULS2. 
Our experiments confirm the significance of hardware-aware programming, especially considering cache memory and avoiding branch mispredictions.\looseness=-1

\section*{Acknowledgement}	
The work was supported by the Polish National Science Center upon decision DEC-2015/17/B/ST6/01890.

\bibliographystyle{splncs03}
\bibliography{ref}

\end{document}